\documentclass[aps,prl,twocolumn,superscriptaddress,oneside,floatfix,amsmath,showpacs,amssymb]{revtex4}
\usepackage{graphicx}
\usepackage{dcolumn}
\usepackage{bm}

\begin{document}

\newcommand{\Y}{YBa$_2$Cu$_3$O$_y$}
\newcommand{\ie}{\textit{i.e.}}
\newcommand{\eg}{\textit{e.g.}}
\newcommand{\etal}{\textit{et al.}}


\title{Onset of a boson mode at the superconducting critical point of underdoped \Y}


\author{Nicolas~Doiron-Leyraud}
\affiliation{D\'epartement de physique and RQMP, Universit\'e de Sherbrooke, Sherbrooke, Canada}

\author{Mike~Sutherland}
\affiliation{Cavendish Laboratory, University of Cambridge, Cambridge, UK}

\author{S.Y.~Li}
\affiliation{D\'epartement de physique and RQMP, Universit\'e de Sherbrooke, Sherbrooke, Canada}

\author{Louis~Taillefer}
\email{Louis.Taillefer@USherbrooke.ca}
\affiliation{D\'epartement de physique and RQMP, Universit\'e de Sherbrooke, Sherbrooke, Canada}
\affiliation{Canadian Institute for Advanced Research, Toronto, Canada}

\author{Ruixing~Liang}
\affiliation{Department of Physics and Astronomy, University of British Columbia, Vancouver, Canada}
\affiliation{Canadian Institute for Advanced Research, Toronto, Canada}

\author{D.A.~Bonn}
\affiliation{Department of Physics and Astronomy, University of British Columbia, Vancouver, Canada}
\affiliation{Canadian Institute for Advanced Research, Toronto, Canada}

\author{W.N.~Hardy}
\affiliation{Department of Physics and Astronomy, University of British Columbia, Vancouver, Canada}
\affiliation{Canadian Institute for Advanced Research, Toronto, Canada}

\date{\today}


\begin{abstract}

The thermal conductivity $\kappa$ of underdoped \Y~was measured in the $T \to 0$ limit as a function of hole concentration $p$ across the superconducting critical point at $p_{SC} \equiv 5.0$~\%. The evolution of bosonic and fermionic contributions to $\kappa$ was tracked as the doping level evolved continuously in each of our samples. For $p \leqslant p_{SC}$, we observe a $T^3$ component in $\kappa$ which we attribute to the boson excitations of a phase with long-range spin or charge order. Fermionic transport, observed as a $T$-linear term in $\kappa$ which persists unaltered through $p_{SC}$, violates the Wiedemann-Franz law, since the electrical resistivity varies as log($1/T$) and grows with decreasing $p$.

\end{abstract}

\maketitle


The nature of the ground state near the point where superconductivity vanishes in underdoped cuprates is of fundamental interest. While a pseudogap and strong phase fluctuations are well established properties of this state, a number of key issues remain open. Is there an order parameter? If so, what symmetry is broken? What are the elementary excitations? Theoretical models addressing these questions involve, for example, the valence bond solid \cite{Senthil}, $d$-density wave order \cite{Nayak}, stripe order \cite{Kivelson}, or fluctuating vortex-antivortex excitations \cite{FranzTesanovic,Herbut}. Recent measurements, on the other hand, show that magnetism can be present up to the point where $T_c$ vanishes \cite{Sanna,Stock}, or that electrons can form a charge-ordered state \cite{Lupien}.

In this Letter, we report a study of the evolution of heat and charge transport across the superconducting transition at $p_{SC}$ in underdoped \Y. In this cuprate, the carrier concentration $p$ in the CuO$_2$ planes is determined by the average oxygen content and the degree to which the dopant atoms are ordered in the chain layers \cite{Veal}. Oxygen ordering occurs at room temperature and the resulting \textit{continuous} change of hole doping with time, called ``time doping'' offers a unique opportunity to track the evolution of \textit{one single specimen} as a function of $p$ without the uncertainty related to changes in cation disorder, or sample and contact characteristics \cite{Broun}.

The thermal conductivity $\kappa$ of any solid may be expressed as the sum of contributions from fermions, bosons (other than phonons), and phonons, respectively:
\begin{equation}
\kappa = \kappa_f + \kappa_b + \kappa_p = aT + \beta T^3 + bT^\alpha~.
\label{eqn:kappa}
\end{equation}
In the $T \to 0$ limit where all inelastic scattering vanishes, the mean free path $\ell$ becomes independent of temperature and frequency and each contribution acquires its distinct asymptotic $T$ dependence, reflecting that of the specific heat $c(T)$ through $\kappa = \frac{1}{3}cv\ell$, where $v$ is the carrier velocity. One exception, as we shall discuss, is the phonon exponent $\alpha$ whose value typically falls below 3 due to specular reflection at the sample boundaries.

In practice, it can be difficult to distinguish $\kappa_b$ from $\kappa_p$ unless one can be ``tuned'' while the other remains constant. This was shown in a recent study of undoped Nd$_2$CuO$_4$ \cite{Li} where heat transport by ballistic magnons, revealed by a large $T^3$ contribution to $\kappa$, could be ``switched on'' by field-tuning across a spin-flop transition. Here, time doping is the tuning method and, like in Nd$_2$CuO$_4$, we have identified a sizable $T^3$ contribution to the thermal conductivity of \Y~that ``switches on'' when $p$ is lowered below $p_{SC}$. We associate this term with non-phononic boson excitations characteristic of the non-superconducting phase.

Heat and charge transport by fermions was also observed. This follows the work of Sutherland \etal~\cite{Sutherland2} and Sun \etal~\cite{Sun}. In the former, a finite and doping-independent $\kappa_f$ due to fermionic excitations was reported for $p$ on both sides of $p_{SC}$. In contrast, for a $y = 6.35$ sample Sun \etal~found $\kappa_f = 0$ and a log($1/T$) dependence of the electrical resistivity, suggesting an insulating state. Here, we measured both properties on \textit{one single specimen} as its doping evolves and found that $\kappa_f$ indeed remains finite and constant close to, and across, $p_{SC}$. Since we also observe a log($1/T$) electrical resistivity below $p_{SC}$, we conclude that at this doping the Wiedemann-Franz law is violated in \Y.

\begin{figure}
\centering
\includegraphics[scale=0.54]{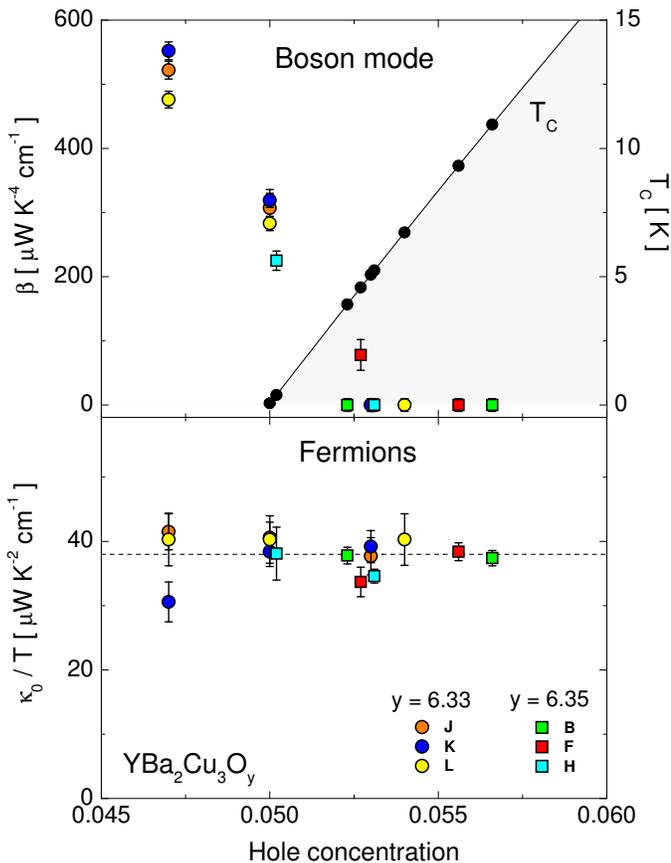}
\caption{Boson and fermion contributions to the thermal conductivity of \Y~vs hole concentration $p$. \textit{Upper panel:} coefficient $\beta$ of bosonic conductivity, $\kappa_b = \beta T^3$. The full line and black dots are the resistive $T_c(p)$ (Note: the bulk $T_c$ lies somewhat lower). For $p \geqslant 5.0$~\%, a finite resistive $T_c$ is observed, from which $p$ is determined \cite{formula}. For $p <$~5.0~\%, no $T_c$ is observed and we use a method described in the text to estimate $p$. \textit{Lower panel:} coefficient $a \equiv \kappa_0/T$ of the fermionic conductivity, $\kappa_f = a T$. Error bars for $\beta$ and $\kappa_0/T$ reflect the statistical uncertainty in the fit. When comparing data from different samples, the uncertainty on the geometric factor, typically between 10 and 20 \%, must be added.}
\label{fig:bosonfermion}
\end{figure}

Ten high-purity single crystals of \Y~were measured, with oxygen content $y = 6.0$ (sample D), $y = 6.31$ (sample O), $y = 6.32$ (samples M, N), $y = 6.33$ (samples J, K, L) and 6.35 (samples B, F, H) (for details, see \cite{Liang}). Sample D is undoped and its hole concentration $p$ is zero. In their oxygen-disordered state, the 6.33 samples were not superconducting, but after 2 (29) days of time doping a $T_c$ of 0.1 (6.0) K was observed. For those and the 6.35 samples, $p$ is determined using the known dependence of $T_c$ on $p$ \cite{formula} and may be read off Fig.~\ref{fig:bosonfermion}. For example, $p$ in sample L evolved in the sequence 4.7, 5.0, and 5.4~\% after 0.25, 2, and 29 days, respectively. Since $p$ varies logarithmically with time \cite{Veal}, we can find the initial hole concentration of our 6.33 samples by extrapolating backward in time. As for the 6.31 and 6.32 samples, the lack of superconducting $T_c$ prohibits a precise determination of $p$, whose initial value we loosely estimate to be near 4 \%. $T_c$ is defined as the temperature where the resistivity becomes zero. The bulk $T_c$ is typically $\simeq 1$~K lower, being the point where the susceptibility drop is complete. No sample was detwinned.

Measurements were made using a four-contact AC method for electrical resistivity and a one-heater-two-thermometer technique for thermal conductivity \cite{Sutherland1}. Great care was taken to ensure a uniform current distribution: very thin samples were used (4--10 $\mu$m thick, 1000--2000 $\mu$m long) with contacts made by evaporating and diffusing gold pads on the sides of each sample.

\begin{figure}
\centering
\includegraphics[scale=0.57]{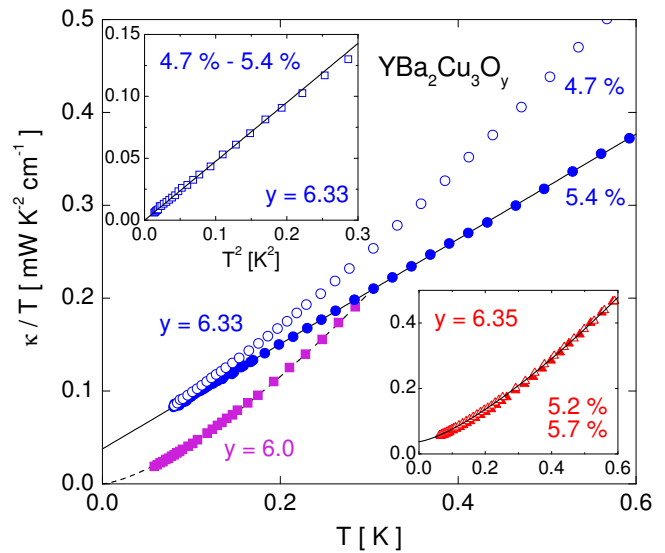}
\caption{Temperature dependence of $\kappa/T$ for three crystals of \Y~with different oxygen content $y$. \textit{Main panel:} $y=6.33$ (sample L) at $p=4.7$~\% (open circles) and 5.4~\% (full circles), with a power-law fit to $\kappa/T = a + bT^{\alpha-1}$ (solid line), giving $\alpha=2.0$; $y=6.0$ ($p=0$) (full squares), with the same fit (dashed line), giving $\alpha=2.4$ ($\kappa/T$ is limited to $T <$~0.3~K and scaled by 1/4 for clarity). \textit{Upper inset:} difference between $\kappa/T$(4.7\%) and $\kappa/T$(5.4\%), vs $T^2$, with a linear fit. \textit{Lower inset:} $y=6.35$ (sample B) at $p=5.2$~\% (open triangles) and 5.7~\% (full triangles), with a fit to the 5.2\% curve that yields $\alpha = 2.4$.}
\label{fig:kappa}
\end{figure}

The thermal conductivity of samples L and B is shown in Fig.~\ref{fig:kappa}, plotted as $\kappa/T$ vs $T$. One can identify three contributions: 1) a $T^3$ term, due to bosons, present below $p = 5.3$~\% but not above, revealed by plotting the difference between successive anneals (upper inset); 2) a residual linear term $\kappa_0/T$, due to fermions, present in both samples; and 3) a phonon background which goes as $T^{\alpha}$ with $\alpha < 3$. We discuss each contribution in detail.


\textit{Bosons}. The data in Fig.~\ref{fig:kappa} shows a generic feature of 6.33 samples: $\kappa$ drops as $p$ increases, by as much as 30~\% at 0.6~K. When plotted vs $T^2$ (upper inset), the difference $\kappa/T(4.7\%) - \kappa/T(5.4\%)$ is linear below 450 mK and a fit to $\Delta\kappa/T = \Delta a + \beta T^2$ gives $\beta = 476$~$\mu$W~K$^{-4}$~cm$^{-1}$ and a negligible intercept ($\Delta a = 0$). A fit to the intermediate difference, $\kappa/T(5.0\%) - \kappa/T(5.4\%)$, yields a similar result, with $\beta = 283$~$\mu$W~K$^{-4}$~cm$^{-1}$ and $\Delta a = 0$. Measurements on more underdoped samples, with $y = 6.31$ and 6.32, show that this $T^3$ contribution to $\kappa$ persists at lower doping. In contrast, however, this term is no longer present above 5.2~\%. This is shown in the lower inset of Fig.~\ref{fig:kappa}, where sample B exhibits negligible change in $\kappa$ as its doping varies from 5.2 to 5.7~\%. Collecting all samples at all dopings in Fig.~\ref{fig:bosonfermion} reveals a systematic trend as $p$ increases: $\beta$ decreases roughly linearly until 5.2~\%, at which point it stops evolving. We thus conclude that $\beta = 0$ for $p >$~5.2~\%. Note that the difference between 5.0~\%, where $T_c$ starts to grow, and 5.2~\%, where $\beta$ extrapolates to zero, is within the combined uncertainty in the bulk $T_c$ and the sampling density.

A pure $T^3$ term in $c(T)$ as $T \to 0$ is the standard signature of bosonic excitations with finite dispersion in all three directions,~\ie, as in any phase with long-range order. For this to be reflected in $\kappa(T)$, the mean free path needs to be independent of $T$, as in the ballistic regime where it is limited only by sample boundaries. In an anisotropic system like cuprates, one can expect $\kappa \sim L T^3 / v_{\perp} v$, where $L$ is the sample size and $v$ ($v_{\perp}$) is the in-plane (out-of-plane) acoustic boson velocity. Such ballistic acoustic bosons offer the most natural interpretation for the $\beta T^3$ term in $\kappa$ observed in \Y.


The appearance of a new boson mode below $p_{SC}$ suggests that long range spin or charge order sets in where superconductivity vanishes. In the case of spin order, the $\beta T^3$ term is ascribed to the magnons of an antiferromagnetic state that develops below $p_{SC}$. This scenario is motivated in part by the striking similarity with the well-understood antiferromagnet Nd$_2$CuO$_4$ \cite{Li}: 1) the ballistic regime where $\kappa_b \propto T^3$ is nearly the same in both cuprates, up to around 0.5~K, and 2) the value of $\beta$ is comparable when considering the difference in sample size. Indeed, although $\beta$ in Nd$_2$CuO$_4$ is 30 times larger than the value in \Y~at 4.7~\%, the three 6.33 samples are 10-15 times thinner. In further support of this scenario, recent neutron scattering measurements on \Y~show the in-plane spin correlation length growing rapidly as $p \to p_{SC}$ (from above) \cite{Buyers}, and $\mu$SR measurements show that the N\'eel temperature of low-doped \Y~declines rapidly as $p \to 0.05$ (from below) \cite{Sanna}. In order to transport heat as $T \to 0$, we note that magnons would have to be gapless at the critical point.

Another scenario is a transition to a Cooper pair Wigner crystal (PWC). Phase fluctuations of the $d$-wave superconducting order parameter can lead to such order \cite{Tesanovic} and recent STM experiments on cuprates (\eg,~\cite{Lupien}) have been interpreted as supporting evidence. Pereg-Barnea and Franz \cite{Franz} recently showed that a PWC has gapless transverse modes of vibration that contribute a $T^3$ term to $\kappa$, with a coefficient they estimate to be in broad agreement with the experimental value of $\beta$ reported here. We note that the fermion sector of the same theory (\cite{Tesanovic,Franz} and references therein) is qualitatively consistent with other aspects of our study, as we shall see below, in the sense that charge is localized in the PWC, but spin can remain mobile, in the form of neutral fermionic spin excitations with a Dirac spectrum.


\textit{Fermions}. The fermionic contribution to $\kappa$ is obtained from the $T=0$ intercept $\kappa_0/T$. For sample L at 5.4~\%, a power-law fit to $\kappa/T = a + bT^{\alpha-1}$ up to 0.6~K yields $a \equiv\kappa_0/T$ = 40~$\pm$~1~$\mu$W~K$^{-2}$~cm$^{-1}$, with $\alpha = 2.0$. As already mentioned, the difference $\Delta\kappa/T$ between $\kappa/T$ at 4.7~\% and $\kappa/T$ at either 5.0 or 5.4~\% yields a negligible intercept, showing that $\kappa_0/T$ for L does not change with doping upon crossing $p_{SC}$. In Fig.~\ref{fig:bosonfermion}, this and the data for five other samples are shown, obtained using the same procedure: a power-law fit for $p \geqslant 5.3$~\% and a $T^2$ fit to the difference for $p < 5.3$~\%. It is remarkable that all six samples measured at $p \simeq 5.3 \%$ gave values within 10~\% of each other: $\kappa_0/T$ = 38 $\pm$ 3~$\mu$W~K$^{-2}$~cm$^{-1}$. This highly reproducible finding confirms the presence of fermionic excitations below $p_{SC}$, as reported recently \cite{Sutherland2}. A magnetic field up to 10 T has negligible effect on $\kappa$ for all samples and dopings.

The fact that $\kappa_0/T$ is nearly constant across $p_{SC}$ suggests that the spectrum of nodal quasiparticles associated with the $d$-wave superconducting gap does not change when superconducting order vanishes. This is what one might expect of a ``nodal metal'', a state characterized in recent angle-resolved photoemission spectroscopy experiments \cite{Kanigel}. Measurements on samples with $y = 6.31$ and 6.32 show that $\kappa_0/T$ eventually becomes negligible as $p$ approaches 4~\%, so that fairly soon this nodal metal phase either comes to an end (\eg, through the appearance of a gap at the nodes) or its fermions become localized.

Our observation of a crossover in $\kappa_0/T$, from vanishing near 4~\% to non-vanishing at 4.7~\% (or below), is qualitatively consistent with the results of Sun \etal~\cite{Sun}, who put the crossover between their more underdoped sample, with $p \simeq 4.5$~\%, and the next one up, with $p \simeq 5.5$~\% ($T_c = 7$~K). They conclude that the crossover occurs halfway, {\it i.e.}, at 5~\%. With the greater resolution conferred by time doping a single sample across $p_{SC}$, we can show unambiguously, as already done in \cite{Sutherland2}, that there is a region of the phase diagram with delocalized fermions and no superconductivity. Note that beyond this issue, there appears to be some discrepancy between their data and ours. While both yield the same value of $\kappa_0/T$ at $p \simeq 6$~\%, theirs gives a significantly lower value at $p \simeq 5.5$~\%. We can only speculate that this ``metal--insulator'' crossover is pushed to slightly higher $p$ in their samples because of a higher impurity level (coming from growth in zirconia vs barium-zirconate crucibles). 


\begin{figure}[t]
\centering
\includegraphics[scale=0.57]{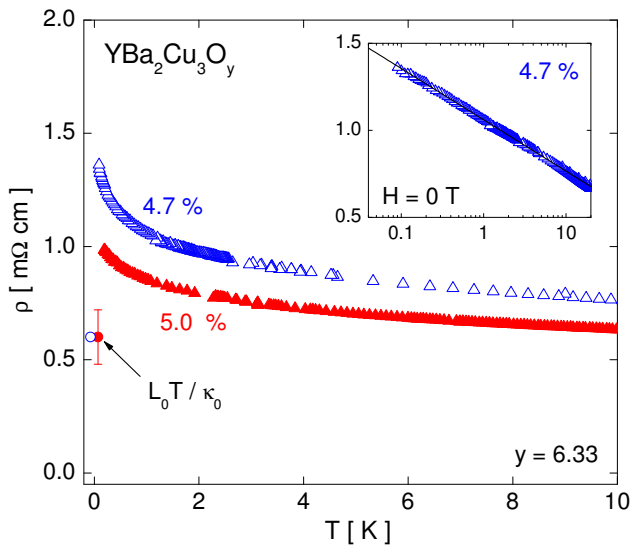}
\caption{Electrical resistivity $\rho$ of non-superconducting YBa$_2$Cu$_3$O$_{6.33}$ at $p=4.7$ (open triangles) and 5.0~\% (full triangles) in $H = 0$ and 10~T, respectively. The corresponding fermionic thermal resistivity is plotted in the same units, as $L_0 T/\kappa_0$ (open and closed circles). \textit{Inset:} $\rho(T)$ at $p=4.7$~\% ($H$~=~0~T) on a logarithmic scale, with linear fit.}
\label{fig:resist}
\end{figure}


Our observation of fermionic heat conduction begs the question as to what happens to charge conduction. Fig.~\ref{fig:resist} shows the normal state electrical resistivity $\rho$ of sample L at 4.7 and 5.0~\% in $H = 0$ and 10~T, respectively. We see that the electrical resistivity drops with increasing doping (at all $T$), as expected from the increase in carrier density, and that it exhibits a perfect log$(1/T)$ behaviour (see inset), over two decades in $T$. The latter shows a clean (and strong) yet unusual tendency towards charge localization, as already reported by Sun \etal~\cite{Sun}.

The measured electrical resistivity is in sharp contrast with the corresponding fermionic heat transport, plotted in Fig.~\ref{fig:resist} in units of resistivity, \ie, $\rho = L_0 T/\kappa_0$, where $L_0 = (\pi^2/3)({\rm k_B}/e)^2$. The significant difference in magnitude reveals an unambiguous violation of the Wiedemann-Franz law (WFL), $\kappa/T = L_0 / \rho$ as $T \to 0$. Beyond the quantitative discrepancy, it is the qualitative trend that is compelling: in the same sample, $\rho$ changes with $p$ while $\kappa_0/T$ remains constant. This is consistent with nodal (and neutral) fermions governed by a Dirac spectrum unrelated to the carrier density, much like $d$-wave quasiparticles in the superconducting state \cite{Sutherland1}.

In the other two instances of WFL violation, observed in Pr$_{2-x}$Ce$_x$CuO$_{4-\delta}$ (PCCO) at optimal doping ($x=0.15$) \cite{Hill} (see \cite{Smith}) and in slightly underdoped Bi$_{2+x}$Sr$_{2-x}$CuO$_{6+\delta}$ (Bi-2201) \cite{Proust}, $\kappa_0/T$ was also found to exceed $L_0 / \rho$. Given the widely different magnitudes and $T$-dependencies of the electrical resistivity in the three cases, more work is needed to elucidate the cause of the violation, and determine whether it is the proximity to a metal-insulator crossover that is relevant, as proposed by Proust \etal~for Bi-2201 \cite{Proust}, or rather the proximity to a magnetic quantum critical point, such as that reported for PCCO at $x=0.16$ \cite{Dagan}.


\textit{Phonons}. Unlike the fermion and boson parts, $\kappa_f$ and $\kappa_b$, whose magnitudes are highly reproducible for the six samples, the phonon background $\kappa_p = \kappa - \kappa_f - \kappa_b = b T^{\alpha}$ is sample dependent (but constant as a function of $p$ for each specimen), with $\alpha$ ranging from 1.8 to 2.5. This is consistent with $\kappa_p$ being dominated by boundary scattering and specular reflection, both dependent on surface dimensions and quality \cite{Sutherland1}. Indeed, $\kappa_p$ in cuprates never goes as $T^3$~\cite{specularboson}. Instead, the existing empirical evidence shows that $\kappa_p = b T^{\alpha}$, with 2~$\lesssim \alpha <$~3, \cite{Sutherland1,Hawthorn,Li}. The most convincing illustration of this sub-$T^3$ dependence comes from data on \textit{undoped} cuprates, insulators in which phonons are the only carriers of heat (since magnons are gapped). In Fig.~\ref{fig:kappa} we show the thermal conductivity data for a typical crystal of YBa$_2$Cu$_3$O$_{6.0}$ (undoped \Y) along with the corresponding fit to $\kappa/T = a + b T^{\alpha-1}$ which describes the data well from 50 to 500 mK, and yields $a \simeq 0$ and $\alpha = 2.4$.


In summary, by carefully tuning the hole concentration $p$ of underdoped \Y~crystals \textit{in-situ} through the superconducting critical point at $p_{SC}$, we have identified the low-lying excitations that characterize the superconductor-insulator transition, from heat transport at $T \to 0$. A novel boson excitation is seen to emerge below $p_{SC}$, most likely the boson of a phase with long-range spin or charge order. Possible interpretations include the magnons of an antiferromagnetic phase or the vibrational modes of a Cooper pair Wigner crystal. On the other hand, the fermionic component of heat conduction persists unchanged through $p_{SC}$, even as the charge conduction decreases and localizes. This suggests that fermion excitations with nodal (and neutral) character survive outside the superconducting phase, at least in the vicinity of the critical point. It is interesting to consider the evolution of charge transport below $p_{SC}$ in relation to the superfluid density above $p_{SC}$ \cite{Broun2}.

We thank M. Franz, A. Chernyshev, D. Broun, R. Hill, and W. Buyers for useful discussions. This work was supported by NSERC (Canada), FQRNT (Qu\'ebec), a Canada Research Chair (L.T.), and the Canadian Institute for Advanced Research.



\begin{references}
\bibitem{specularboson} Note that specular reflection appears to be ineffective for the bosons, just as it was for magnons in Nd$_2$CuO$_4$ \cite{Li}.
\end{references}
\end{document}